\documentclass[thmsa,notitlepage,11pt]{article}
\usepackage{sw20jart}

\input tcilatex
\begin{document}

\title{IV Estimation of Panel Data Tobit Models with Normal Errors\thanks{%
Financial support from the Alfred P. Sloan Foundation and from the National
Science Foundation are gratefully acknowledged, as are comments from
numerous seminar participants. Kenneth Y. Chay, Luojia Hu and Ekaterini
Kyriazidou gave useful comments on an earlier draft.}.}
\author{Bo E. Honor\'{e}\thanks{%
Mailing address: Department of Economics, Princeton University, Princeton,
NJ 08544-1021. Phone (609) 258-4014. Fax: (609) 258-5561. Email:
honore@Princeton.edu. }}
\date{First complete draft: April 1996. This version: May 1998.}
\maketitle

\begin{abstract}
Amemiya (1973) proposed a ``consistent initial estimator'' for the
parameters in a censored regression model with normal errors. This paper
demonstrates that a similar approach can be used to construct moment
conditions for fixed--effects versions of the model considered by Amemiya.
This result suggests estimators for models that have not previously been
considered.
\end{abstract}

\section{A Moment for The Truncated Bivariate Normal Distribution.}

The approach taken in this paper is motivated by Amemiya's ``initial
consistent estimator'' (Amemiya 1973). That estimator uses the observation
that if 
\begin{equation}
y_i=\max \left\{ 0,x_i^{\prime }\beta +\varepsilon _i\right\}  \label{cr}
\end{equation}
with $\varepsilon _i\sim N(0,\sigma ^2)$ (independent of $x_i$), then 
\[
E\left[ y_i^2|y_i>0,x_i\right] =E\left[ y_i|y_i>0,x_i\right] x_i^{\prime
}\beta +\sigma ^2 
\]
which implies that 
\begin{equation}
y_i^2=y_ix_i^{\prime }\beta +\sigma ^2+\xi _i,\qquad E\left[ \xi
_i|y_i>0,x_i\right] =0.  \label{iv}
\end{equation}
One can therefore estimate $\beta $ and $\sigma ^2$ by applying instrumental
variables to $\left( \ref{iv}\right) $, using only observations for which $%
y_i>0,$ and using functions of $x_i$ as instruments.

The contribution of this paper is to demonstrate that Amemiya's result can
be generalized to panel data versions of (\ref{cr}) with
individual--specific effects, 
\begin{equation}
y_{it}=\max \left\{ 0,x_{it}^{\prime }\beta +\alpha _i+\varepsilon
_{it}\right\} ,  \label{pd}
\end{equation}
where $\left\{ \varepsilon _{it}\right\} _{t=1}^T$ is normally distributed
and independent of $\left( \left\{ x_{it}\right\} _{t=1}^T,\alpha _i\right) $%
. Throughout, we will assume random sampling across individuals, $i$.

The model in (\ref{pd}) was used by Heckman and MaCurdy's (1980) to study
female labor supply. In their model, $\alpha _i$ is a function of, among
other things, the Lagrange multiplier associated with an individual's budget
constraint, which in turn ``is a messy, but scalar, function of wage rates,
interest rates,...'' This implies that it is not reasonable to try to model $%
\alpha _i$ as a function of the explanatory variables, and they therefore
treat $\left\{ \alpha _i\right\} _{i=1}^n$ as a set of parameters to be
estimated. This estimation procedure is justified asymptotically as $%
T\rightarrow \infty $. Honor\'{e} (1992) proposed estimators\footnote{%
A recent paper by Honor\'{e} and Kyriazidou (1998) shows that it is possible
to modify Honor\'{e}'s (1992) estimator in such a way that the
exchangeability assumption can be replaced by a stationarity assumption.}
for $\beta $ that are justified asymptotically as $N\rightarrow \infty $
(with $T$ fixed) under the exchangeability assumption that for any two time
periods $t$ and $s$, $\left( \varepsilon _{it},\varepsilon _{is}\right) $ is
distributed like $\left( \varepsilon _{is},\varepsilon _{it}\right) $
conditional on $\left( x_{it}^{\prime },x_{is}^{\prime },\alpha _i\right) $.
Here Amemiya's approach will be used to construct moment conditions (and
hence, implictly, instrumental variables estimators) that can be applied to
versions of the fixed effects censored regression model that assume normal
errors, but are more general than the model considered in Honor\'{e} (1992)
in other dimensions.

Before proceeding to develop versions of $\left( \ref{iv}\right) $ that can
be applied to panel data, we note that it is possible to extend $\left( \ref
{iv}\right) $ to higher moments. A simple integration--by--parts argument
can be used to show that if $U\sim N(\mu ,\sigma ^2)$ then for any $k\geq 1$%
\[
E\left[ U^{k+1}|U>0\right] =\mu E\left[ U^k|U>0\right] +\sigma ^2kE\left[
U^{k-1}|U>0\right] 
\]
Applying this to the censored regression model $y_i=\max \left\{
0,x_i^{\prime }\beta +\varepsilon _i\right\} ,$ with $\varepsilon _i\sim
N(0,\sigma ^2),$ and $\varepsilon _i$ independent of $x_i$, yields the
moment conditions 
\[
E\left[ y_i^{k+1}-y_i^kx_i^{\prime }\beta -\sigma
^2ky_i^{k-1}|y_i>0,x_i\right] =0 
\]
so $\beta $ and $\sigma ^2$ can be estimated by applying instrumental
variables estimation to the equation 
\begin{equation}
y_i^{k+1}=\left( y_i^kx_i^{\prime }\right) \beta +\left( ky_i^{k-1}\right)
\sigma ^2+\xi _i,  \label{am2}
\end{equation}
using only the positive $y_i$'s and using functions of $x_i$ as instruments.
For $k=1$ this is equation $\left( \ref{iv}\right) $ used by Amemiya (1973).

For the panel data censored regression model, (\ref{pd}), the equation
underlying Amemiya's estimator becomes 
\begin{equation}
E\left[ y_{it}^2-y_{it}\left( x_{it}^{\prime }\beta +\alpha _i\right)
-\sigma _t^2|x_{it},\alpha _i,y_{it}>0\right] =0  \label{m1}
\end{equation}
where $\sigma _t^2$ denotes the variance of $\varepsilon _{it}$. If $%
\varepsilon _{it}$ and $\varepsilon _{is}$ are \textit{independent} of each
other, and of $\left( x_{it},x_{is},\alpha _i\right) $, then (\ref{m1}) also
holds if we also condition on $y_{is}$. This, in turn, implies that 
\[
E\left[ y_{it}^2y_{is}-y_{it}y_{is}\left( x_{it}^{\prime }\beta +\alpha
_i\right) -y_{is}\sigma _t^2|x_{it},x_{is},\alpha _i,y_{is},y_{it}>0\right]
=0 
\]
and therefore by the law of iterated expectations 
\[
E[y_{it}^2y_{is}-y_{it}y_{is}\left( x_{it}^{\prime }\beta +\alpha _i\right)
-y_{is}\sigma _t^2|x_{it},x_{is},\alpha _i,y_{is}>0,y_{it}>0]=0. 
\]
Reversing $t$ and $s,$ taking differences, and integrating out $\alpha _i$,
yields 
\begin{eqnarray}
&&E[(y_{it}^2y_{is}-y_{is}^2y_{it})-y_{it}y_{is}(x_{it}^{\prime
}-x_{is}^{\prime })\beta  \label{hm1} \\
&&-\left( y_{is}\sigma _t^2-y_{it}^{}\sigma _s^2\right)
|x_{it},x_{is},y_{is}>0,y_{it}>0]=0.  \nonumber
\end{eqnarray}
or equivalently 
\begin{equation}
(y_{it}^2y_{is}-y_{is}^2y_{it})=y_{it}y_{is}(x_{it}^{\prime }-x_{is}^{\prime
})\beta +y_{is}\sigma _t^2-y_{it}\sigma _s^2+\xi _{ist},  \label{hm1a}
\end{equation}
with $E[\xi _{ist}|x_{it},x_{is},y_{is}^{}>,y_{it}^{}>0]$. This implies that 
$\beta $, $\sigma _t^2$ and $\sigma _s^2$ can be estimated by applying
instrumental variables to (\ref{hm1a}), using functions of the $x$'s as
instruments. The statistical properties of such an estimator then follow by
standard instrumental variables arguments.

The derivation leading to (\ref{hm1}) assumes that the errors are
independently distributed over time. The following proposition (which is
proved in the appendix) will allow us to derive moment conditions which
exploit higher moments as in (\ref{am2}) and which can be applied when the
errors have a joint normal distribution with arbitrary dependence.

\begin{proposition}
\label{prop1}If 
\[
\left( 
\begin{array}{c}
U_1 \\ 
U_2
\end{array}
\right) \sim N\left( \left( 
\begin{array}{c}
\mu _1 \\ 
\mu _2
\end{array}
\right) ,\left( 
\begin{array}{cc}
\sigma _1^2 & \sigma _{12} \\ 
\sigma _{12} & \sigma _2^2
\end{array}
\right) \right) 
\]
then for $k\geq 1$ and $m\geq 1$, 
\begin{eqnarray*}
&&E\left[ \left. U_i^{k+1}U_j^m-U_i^kU_j^{m+1}\right| U_i>0,U_j>0\right] \\
&=&\left( \mu _i-\mu _j\right) E\left[ \left. U_i^kU_j^m\right|
U_i>0,U_j>0\right] \\
&&+\left( \sigma _i^2-\sigma _{ij}\right) kE\left[ \left.
U_i^{k-1}U_j^m\right| U_i>0,U_j>0\right] \\
&&-\left( \sigma _j^2-\sigma _{ij}\right) mE\left[ \left.
U_i^kU_j^{m-1}\right| U_i>0,U_j>0\right]
\end{eqnarray*}
\end{proposition}

\section{Moment Conditions for Truncated and Censored Regression Models.}

Proposition 1.1 can be used to construct moment conditions for panel data
censored or truncated regression models under a variety of assumptions,
provided that one is willing to assume that the errors are normally
distributed conditional on the regressors of the model and conditional on
the fixed effects.

\subsection{Non-stationarity.}

It is convenient to rewrite the censored regression model\footnote{%
With truncation $\left( y_{it},x_{it}\right) $, are drawn from (\ref{ystar})
conditional of $y_{it}^{*}>0$. The methods described here can be applied to
truncated regression models as well as to censored regression models.} in
equation(\ref{pd}) in terms of an underlying latent variable, $y_{it}^{*}$, 
\begin{eqnarray}
y_{it}^{*} &=&x_{it}^{\prime }\beta +\alpha _i+\varepsilon _{it},\qquad
t=1,\ldots ,T,\qquad i=1,\ldots ,N  \label{ystar} \\
y_{it} &=&\max \left\{ 0,y_{it}^{*}\right\}  \nonumber
\end{eqnarray}
In this subsection we will assume that $\left( \varepsilon _{i1},\ldots
,\varepsilon _{iT}\right) $ is normally distributed and independent of $%
\left( \alpha _i,x_{i1},\ldots ,x_{iT}\right) $ with var$\left[ \varepsilon
_{it}\right] =\sigma _t^2$ and cov$\left( \varepsilon _{it},\varepsilon
_{is}\right) =\sigma _{t,s}$. The generalization here, relatively to
Honor\'{e} (1992), is that it is not necessary to assume that $\left\{
\varepsilon _{it}\right\} $ is stationary. In some contexts, this can be
important. For example, Chay (1995) considered panel data censored
regression models for wages\footnote{%
In his application, the censoring was induced because he used U.S. social
security earnings as his dependent variable. This implies that the
observations are censored from above at the taxable maximum.}. Since it is
fairly well--documented that the distribution of earnings has changed over
time in the U.S., it does not seem reasonable to assume that $\left\{
\varepsilon _{it}\right\} $ is stationary (and Chay consequently took a
random effects approach, rather than a fixed effects approach).

Applying Proposition 1.1 to the conditional distribution of $\left(
y_{it}^{*},y_{is}^{*}\right) $ given $\left( x_{it}^{\prime },x_{is}^{\prime
},\alpha _i\right) $ yields 
\begin{eqnarray}
\left. y_{it}^{*}\right. ^{k+1}\left. y_{is}^{*}\right. ^m-\left.
y_{is}^{*}\right. ^{m+1}\left. y_{it}^{*}\right. ^k &=&\left.
y_{it}^{*}\right. ^k\left. y_{is}^{*}\right. ^m\left( x_{it}^{\prime
}-x_{is}^{\prime }\right) \beta +k\left. y_{it}^{*}\right. ^{k-1}\left.
y_{is}^{*}\right. ^m\left( \sigma _t^2-\sigma _{s,t}\right)  \label{Estm} \\
&&-m\left. y_{is}^{*}\right. ^{m-1}\left. y_{it}^{*}\right. ^k\left( \sigma
_s^2-\sigma _{s,t}\right) +\xi _{ist}  \nonumber
\end{eqnarray}
where 
\[
E\left[ \xi _{ist}|x_{it}^{\prime },x_{is}^{\prime
},y_{it}^{*}>0,y_{is}^{*}>0\right] =0. 
\]

This implies that $\beta $, $\left( \sigma _s^2-\sigma _{t,s}\right) $ and $%
\left( \sigma _t^2-\sigma _{t,s}\right) $ can be estimated by using only
pairs of observations for which $y_{it}>0$ and $y_{is}>0$ and then applying
instrumental variable estimation to (\ref{Estm}) using functions of $\left(
x_{it}^{\prime },x_{is}^{\prime }\right) $ as instruments.

\subsection{Factor loading.}

It is sometimes desirable to allow the individual specific effect to have
different coefficients in different time periods. Such a model has been
discussed in a number of contexts. For example, Heckman (1981) investigates
a random effect discrete choice model, Holtz--Eakin, Newey and Rosen (1988)
a linear autoregressive model, and Chay (1995) a random effects censored
regression model, where (in all cases) the individual specific effect is
multiplied by a time--specific ``factor--loading''. Proposition 1.1 can be
used to construct an estimator of a fixed effect censored regression model
of this kind.

The model is

\begin{equation}
y_{it}^{*}=x_{it}^{\prime }\beta +\rho _t\alpha _i+\varepsilon _{it},\qquad
t=1,\ldots ,T,\qquad i=1,\ldots ,N  \label{ystar_fl}
\end{equation}
where the researcher observes $x_{it}$ and $y_{it}=\max \left\{
0,y_{it}^{*}\right\} $.

Since $\alpha _i$ is not observed, it is not possible to identify the scale
of the factor--loadings, $\rho _1,\rho _2,\ldots \rho _T$. However, for any
two time periods, $t$ and $s$, we have 
\begin{eqnarray}
y_{it}^{*2}y_{is}^{*}-y_{is}^{*2}y_{it}^{*}\rho _t/\rho _s
&=&y_{it}^{*}y_{is}^{*}\left( x_{it}-x_{is}\rho _t/\rho _s\right) ^{\prime
}\beta  \label{Estm_fl} \\
&&-y_{it}^{*}\left( \sigma _s^2\rho _t/\rho _s-\sigma _{t,s}\right)
+y_{is}^{*}\left( \sigma _t^2-\sigma _{t,s}\rho _t/\rho _s\right) +\xi _{ist}
\nonumber
\end{eqnarray}
where, again, 
\[
E\left[ \xi _{ist}|x_{it}^{\prime },x_{is}^{\prime
},y_{it}^{*}>0,y_{is}^{*}>0\right] =0. 
\]
As was the case in the previous subsection, it is also possible to use
information contained in higher moments.

\subsection{Fixed effects in variance.}

The homoskadasticity assumption across individuals is strong. The other
extreme is to assume that the errors are heteroskedastic across
observations, but\textsl{\ i.i.d.} over time, and with different variances
for different individuals. In that case $\left( \text{\ref{Estm}}\right) $
becomes 
\begin{equation}
y_{it}^{*2}y_{is}^{*}-y_{is}^{*2}y_{it}^{*}=y_{it}^{*}y_{is}^{*}\left(
x_{it}-x_{is}\right) ^{\prime }\beta -\left( y_{it}^{*}-y_{is}^{*}\right)
\sigma _i^2+\xi _{ist}  \label{Estm_h1}
\end{equation}
with 
\[
E\left[ \xi _{ist}|x_{it}^{\prime },x_{is}^{\prime
},y_{it}^{*}>0,y_{is}^{*}>0\right] =0. 
\]
Considering also a third time period, $\tau $, we get 
\begin{eqnarray}
&&\left( y_{it}^{*2}y_{is}^{*}-y_{is}^{*2}y_{it}^{*}\right) +\left(
y_{is}^{*2}y_{i\tau }^{*}-y_{i\tau }^{*2}y_{is}^{*}\right) +\left( y_{i\tau
}^{*2}y_{it}^{*}-y_{it}^{*2}y_{i\tau }^{*}\right)  \label{Estm_h2} \\
&=&y_{it}^{*}y_{is}^{*}\left( x_{it}-x_{is}\right) ^{\prime }\beta
+y_{is}^{*}y_{i\tau }^{*}\left( x_{is}-x_{i\tau }\right) ^{\prime }\beta
+y_{i\tau }^{*}y_{it}^{*}\left( x_{i\tau }-x_{it}\right) ^{\prime }\beta 
\nonumber \\
&&-\left( y_{it}^{*}-y_{is}^{*}\right) \sigma _i^2-\left(
y_{is}^{*}-y_{i\tau }^{*}\right) \sigma _i^2-\left( y_{i\tau
}^{*}-y_{it}^{*}\right) \sigma _i^2  \nonumber \\
&&+\xi _{ist}+\xi _{i\tau s}+\xi _{it\tau }  \nonumber \\
&=&\left( y_{it}^{*}y_{is}^{*}\left( x_{it}-x_{is}\right) ^{\prime
}+y_{is}^{*}y_{i\tau }^{*}\left( x_{is}-x_{i\tau }\right) ^{\prime
}+y_{i\tau }^{*}y_{it}^{*}\left( x_{i\tau }-x_{it}\right) ^{\prime }\right)
\beta +\xi _{ist\tau }  \nonumber
\end{eqnarray}
with 
\[
E\left[ \xi _{ist\tau }|x_{it}^{\prime },x_{is}^{\prime },x_{i\tau }^{\prime
},y_{it}^{*}>0,y_{is}^{*}>0,y_{i\tau }^{*}>0\right] =0. 
\]

It is also possible to allow the variance to have time-specific component,
provided that the resulting variance has an additive structure of the form $%
\sigma _{it}^2=\sigma _i^2+\sigma _t^2$. In this case we have 
\begin{eqnarray}
&&\left( y_{it}^{*2}y_{is}^{*}-y_{is}^{*2}y_{it}^{*}\right) +\left(
y_{is}^{*2}y_{i\tau }^{*}-y_{i\tau }^{*2}y_{is}^{*}\right) +\left( y_{i\tau
}^{*2}y_{it}^{*}-y_{it}^{*2}y_{i\tau }^{*}\right)  \label{Estm_h3} \\
&=&y_{it}^{*}y_{is}^{*}\left( x_{it}-x_{is}\right) ^{\prime }\beta
+y_{is}^{*}y_{i\tau }^{*}\left( x_{is}-x_{i\tau }\right) ^{\prime }\beta
+y_{i\tau }^{*}y_{it}^{*}\left( x_{i\tau }-x_{it}\right) ^{\prime }\beta 
\nonumber \\
&&-y_{it}^{*}\left( \sigma _s^2+\sigma _i^2\right) +y_{is}^{*}\left( \sigma
_t^2+\sigma _i^2\right) -y_{is}^{*}\left( \sigma _\tau ^2+\sigma _i^2\right)
\nonumber \\
&&+y_{i\tau }^{*}\left( \sigma _s^2+\sigma _i^2\right) -y_{i\tau }^{*}\left(
\sigma _t^2+\sigma _i^2\right) +y_{it}^{*}\left( \sigma _\tau ^2+\sigma
_i^2\right)  \nonumber \\
&&+\xi _{ist}+\xi _{i\tau s}+\xi _{it\tau }  \nonumber \\
&=&\left( y_{it}^{*}y_{is}^{*}\left( x_{it}-x_{is}\right) ^{\prime
}+y_{is}^{*}y_{i\tau }^{*}\left( x_{is}-x_{i\tau }\right) ^{\prime
}+y_{i\tau }^{*}y_{it}^{*}\left( x_{i\tau }-x_{it}\right) ^{\prime }\right)
\beta  \nonumber \\
&&+\left( y_{i\tau }^{*}-y_{it}^{*}\right) \sigma _s^2+\left(
y_{is}^{*}-y_{i\tau }^{*}\right) \sigma _t^2+\left(
y_{it}^{*}-y_{is}^{*}\right) \sigma _\tau ^2  \nonumber \\
&&+\xi _{ist\tau }  \nonumber
\end{eqnarray}
with 
\[
E\left[ \xi _{ist\tau }|x_{it}^{\prime },x_{is}^{\prime },x_{i\tau }^{\prime
},y_{it}^{*}>0,y_{is}^{*}>0,y_{i\tau }^{*}>0\right] =0. 
\]
As in the previous subsection, additional moment conditions can be obtained
by applying Proposition 1 for other values of $m$ and $k$.

\subsection{Fixed Effects in the Slopes.}

Rather than letting the fixed effect work through the level of the model,
one might be interested in estimating $\beta $ in the model 
\begin{equation}
y_{it}^{*}=x_{it}^{\prime }\beta +z_{it}\alpha _i+\varepsilon _{it},\qquad
t=1,\ldots ,T,\qquad i=1,\ldots ,N  \label{ystar_sl}
\end{equation}
where $z_{it}$ is an observed explanatory variable. This model is in the
spirit of random coefficients models.

In this case, we can apply Proposition 1 to the distribution of\footnote{%
To simplify the derivations, we assume here that $z_{it}>0$ and $z_{is}>0$.} 
$\left( y_{is}^{*}z_{it},y_{it}^{*}z_{is}\right) $, conditional on $\left(
x_{it},x_{is},z_{it},z_{is},\alpha _i\right) $, 
\begin{eqnarray}
y_{it}^{*2}z_{is}^2y_{is}^{*}z_{it}-y_{is}^{*2}z_{it}^2y_{it}^{*}z_{is}
&=&y_{it}^{*}z_{is}y_{is}^{*}z_{it}\left( x_{it}z_{is}-x_{is}z_{it}\right)
^{\prime }\beta -  \label{Estm-sl} \\
&&y_{it}^{*}z_{is}\left( z_{it}^2\sigma _s^2-z_{it}z_{is}\sigma
_{t,s}\right) +y_{is}^{*}z_{it}\left( z_{is}^2\sigma _t^2-z_{it}z_{is}\sigma
_{t,s}\right) +\xi _{ist}  \nonumber
\end{eqnarray}
with 
\[
E\left[ \xi _{ist}|z_{it},z_{is},x_{it}^{\prime },x_{is}^{\prime
},y_{it}^{*}>0,y_{is}^{*}>0\right] =0. 
\]

\section{Concluding Remarks.}

The moment conditions and associated instrumental variables procedures
discussed in this paper all rely on strict exogeneity of the explanatory
variables\footnote{%
Formally, assumptions are made on the error terms, conditional on future
values of the explanatory variales. This is very restrictive, as it rules
out feedback from the current dependent variables to future explanatory
variables.} and on normality of the errors. While these are strong
assumptions, the methods proposed here are the first that can be applied to
the ``fixed--effects'' Tobit models in previous sections.

Wales and Woodland (1980) found that the estimator of the cross--sectional
censored regression model based on (\ref{iv}) is relatively inefficient.
Since the approach here is based on the same insight as (\ref{iv}), one
might worry that estimators based on the Proposition 1.1 will also be
imprecise. The difference between the two cases, however, is that for the
censored regression model with normal errors, there are other estimators to
which the instrumental variables estimator can be compared. In the panel
data settings discussed here, the instrumental variables estimator is the
first, and so far only, available estimator.

Finally, we note that for all the example considered here, one can construct
estimators based on any choice of $k$ and $m$ in Proposition 1.1. Combining
these moment conditions optimally would certainly improve the asymptotic
efficiency of the estimator. We also note that the moment conditions here
are conditional, and that the errors are heteroskedastic. This implies
additional methods for improving the asymptotic efficiency of the estimator.
However, the contribution of this paper is to show that it is\textit{\
possible} to estimated the models considered in the previous section, and
not to re--derive results about optimal generalized method of moments
estimation.

\section{References.}

\begin{description}
\item  Amemiya, T. (1973): ``Regression Analysis when the Dependent Variable
is Truncated Normal,'' Econometrica, 41, pp.~997--1016.

\item  Chay, Kenneth Y. (1995): ``Evaluating the Impact of the 1964 Civil
Rights Act on the Economics Status of Black Men using Censored Longitudinal
Earnings Data,'' unpublished manuscript, Department of Economics, Princeton
University.

\item  Heckman, J. J. (1981): ``Statistical Models for Discrete Panel
Data.'' In McFadden, D. and Manski, C. (eds.) ``Structural Analysis of
Discrete Data with Econometric Applications'', Cambridge, Mass.: MIT Press.

\item  Heckman, J. J. and T. E. MaCurdy (1980): ``A Life Cycle Model of
Female Labour Supply.'' Review of Economic Studies, 47, pp. 47--74.

\item  Holtz--Eakin, D., W. Newey and H. S. Rosen (1988): ``Estimating
Vector Autoregressions with Panel Data.'' Econometrica, 56, pp. 1371-1395.

\item  Honor\'{e}, B.~E.\ (1992): ``Trimmed LAD and Least Squares Estimation
of Truncated and Censored Regression Models with Fixed Effects,''
Econometrica, 60, pp.\ 533--565.

\item  Honor\'{e}, B. E., and E. Kyriazidou (1998): ``Estimation of
Tobit--Type Models with Individual Specific Effects.'' Unpublished.

\item  Wales, T. J., and A. D. Woodland (1980): ``Sample Selectivity and the
Estimation of Labor Supply Functions.'' International Economic Review 21:
437--468.
\end{description}

\section{Appendix: Derivation of Proposition 1.1.}

Using the usual notation, the bivariate normal density, $f\left(
u_1,u_2\right) $ satisfies 
\[
\frac{\partial f\left( u_1,u_2\right) }{\partial u_1}=-\frac 1{1-\rho
^2}\left( \frac 1{\sigma _1^2}u_1-\frac \rho {\sigma _1\sigma _2}u_2-\frac{%
\mu _1}{\sigma _1^2}+\frac{\rho \mu _2}{\sigma _1\sigma _2}\right) f\left(
u_1,u_2\right) 
\]
Therefore 
\begin{eqnarray*}
&&\int_0^\infty ku_1^{k-1}u_2^mf\left( u_1,u_2\right) du_1 \\
&=&\int_0^\infty u_1^ku_2^m\frac 1{1-\rho ^2}\left( \frac 1{\sigma
_1^2}u_1-\frac \rho {\sigma _1\sigma _2}u_2-\frac{\mu _1}{\sigma _1^2}+\frac{%
\rho \mu _2}{\sigma _1\sigma _2}\right) f\left( u_1,u_2\right) du_1 \\
&=&\frac 1{1-\rho ^2}\int_0^\infty \left( \frac 1{\sigma
_1^2}u_1^{k+1}u_2^m-\frac \rho {\sigma _1\sigma _2}u_1^ku_2^{m+1}-\left( 
\frac{\mu _1}{\sigma _1^2}-\frac{\rho \mu _2}{\sigma _1\sigma _2}\right)
u_1^ku_2^m\right) f\left( u_1,u_2\right) du_1
\end{eqnarray*}
or 
\begin{eqnarray*}
&&k\left( 1-\rho ^2\right) \sigma _1^2\sigma _2E\left[ \left.
U_1^{k-1}U_2^m\right| U_1>0,U_2>0\right] \\
&=&E\left[ \left. \sigma _2U_1^{k+1}U_2^m-\rho \sigma
_1U_1^kU_2^{m+1}-\left( \sigma _2\mu _1-\rho \sigma _1\mu _2\right)
U_1^kU_2^m\right| U_1>0,U_2>0\right]
\end{eqnarray*}
multiplying by $\sigma _2$%
\begin{eqnarray}
&&k\left( \sigma _1^2\sigma _2^2-\sigma _{12}^2\right) E\left[ \left.
U_1^{k-1}U_2^m\right| U_1>0,U_2>0\right]  \label{ap1} \\
&=&E\left[ \left. \sigma _2^2U_1^{k+1}U_2^m-\sigma
_{12}U_1^kU_2^{m+1}-\left( \sigma _2^2\mu _1-\sigma _{12}\mu _2\right)
U_1^kU_2^m\right| U_1>0,U_2>0\right]  \nonumber
\end{eqnarray}
Applying the same integration by parts to $\int_0^\infty
mu_1^ku_2^{m-1}f\left( u_1,u_2\right) du_2$, we get 
\begin{eqnarray}
&&m\left( \sigma _1^2\sigma _2^2-\sigma _{12}^2\right) E\left[ \left.
U_1^kU_2^{m-1}\right| U_1>0,U_2>0\right]  \label{ap2} \\
&=&E\left[ \left. \sigma _1^2U_1^kU_2^{m+1}-\sigma
_{12}U_1^{k+1}U_2^m-\left( \sigma _1^2\mu _2-\sigma _{12}\mu _1\right)
U_1^kU_2^m\right| U_1>0,U_2>0\right]  \nonumber
\end{eqnarray}
Multiplying $\left( \text{\ref{ap1}}\right) $ by $\left( \sigma _1^2-\sigma
_{12}\right) $ and $\left( \text{\ref{ap2}}\right) $ by $\left( \sigma
_2^2-\sigma _{12}\right) $, and then subtracting $\left( \text{\ref{ap2}}%
\right) $ from $\left( \text{\ref{ap1}}\right) $ and simplifying the right
hand side yields 
\begin{eqnarray*}
&&k\left( \sigma _1^2-\sigma _{12}\right) \left( \sigma _1^2\sigma
_2^2-\sigma _{12}^2\right) E\left[ \left. U_1^{k-1}U_2^m\right|
U_1>0,U_2>0\right] \\
&&-m\left( \sigma _2^2-\sigma _{12}\right) \left( \sigma _1^2\sigma
_2^2-\sigma _{12}^2\right) E\left[ \left. U_1^kU_2^{m-1}\right|
U_1>0,U_2>0\right] \\
&=&\left[ \left( \sigma _1^2-\sigma _{12}\right) \sigma _2^2+\left( \sigma
_2^2-\sigma _{12}\right) \sigma _{12}\right] E\left[ \left.
U_1^{k+1}U_2^m\right| U_1>0,U_2>0\right] \\
&&-\left[ \left( \sigma _1^2-\sigma _{12}\right) \sigma _{12}+\left( \sigma
_2^2-\sigma _{12}\right) \sigma _1^2\right] E\left[ \left.
U_1^kU_2^{m+1}\right| U_1>0,U_2>0\right] \\
&&-\left[ \left( \sigma _1^2-\sigma _{12}\right) \left( \sigma _2^2\mu
_1-\sigma _{12}\mu _2\right) -\left( \sigma _2^2-\sigma _{12}\right) \left(
\sigma _1^2\mu _2-\sigma _{12}\mu _1\right) \right] E\left[ \left.
U_1^kU_2^m\right| U_1>0,U_2>0\right] \\
&=&\left( \sigma _1^2\sigma _2^2-\sigma _{12}^2\right) E\left[ \left.
U_1^{k+1}U_2^m\right| U_1>0,U_2>0\right] \\
&&-\left( \sigma _1^2\sigma _2^2-\sigma _{12}^2\right) E\left[ \left.
U_1^kU_2^{m+1}\right| U_1>0,U_2>0\right] \\
&&-\left( \sigma _1^2\sigma _2^2-\sigma _{12}^2\right) \left( \mu _1-\mu
_2\right) E\left[ \left. U_1^kU_2^m\right| U_1>0,U_2>0\right]
\end{eqnarray*}
dividing through by $\left( \sigma _1^2\sigma _2^2-\sigma _{12}^2\right) $
then yields the result.

\end{document}